\documentclass[a4paper]{article}
\pdfoutput=1

\usepackage[english]{babel}
\usepackage[utf8]{inputenc}
\usepackage{natbib}
\usepackage{amsmath}
\usepackage{amsthm}
\usepackage{amsfonts}
\usepackage{graphicx}
\usepackage{url}
\usepackage{listings}
\usepackage{array}
\usepackage{verbatim}
\usepackage{longtable}
\usepackage[official]{eurosym}
\usepackage{algorithm}
\usepackage[noend]{algpseudocode}
\usepackage{tikz}
\usetikzlibrary{shapes,arrows, shapes.geometric, shapes.symbols}
\everymath{\displaystyle}

\theoremstyle{definition}

\usepackage[font=scriptsize]{caption}

\usepackage{authblk}

\tikzstyle{every label}= [black]

\tikzstyle{place}=[circle,draw=black,minimum size=6mm, node distance=2cm]
\tikzstyle{transition}=[rectangle,draw=black, minimum size=6mm, node distance=2cm]
\tikzstyle{pre}=[<-,shorten <=1pt,>=stealth',semithick]
\tikzstyle{post}=[->,shorten >=1pt,>=stealth',semithick]

\tikzstyle{block} = [rectangle, draw, fill=gray!50, 
    text width=3cm, text centered, rounded corners, minimum height=4em]
\tikzstyle{line} =  [draw, thick, ->, shorten >=2pt] 
\tikzstyle{cloud} = [draw, ellipse,fill=red!20, node distance=3cm,
    minimum height=2em]
\tikzstyle{print} = [draw, tape, tape bend top=none, fill=gray!50, node distance=3cm
	 text width=5em, minimum height=4em, text centered]
\tikzstyle{decision}= [diamond, aspect=2, draw, fill=gray!50,
                        text width=2.5cm, text centered,
                        inner sep=1pt, rounded corners]

\begin{document}

\title{Empirical analysis of daily cash flow time series and its implications for forecasting}

\author[1]{Francisco Salas-Molina\footnote{Corresponding author. E-mail addresses: \textit{francisco.salas@hifesa.com, jar@iiia.csic.es, joan.serra@telefonica.com, mguillen@ub.edu, martin@bigml.com, }}}
\author[2]{Juan A. Rodr\'iguez-Aguilar}
\author[3]{Joan Serr\`a}
\author[4]{Montserrat Guillen}
\author[5]{Francisco J. Martin}

\affil[1]{Hilaturas Ferre, S.A., Les Molines, 2, 03450 Banyeres de Mariola, Alicante, Spain}
\affil[2]{IIIA-CSIC, Campus UAB, 08913 Cerdanyola, Catalonia, Spain}
\affil[3]{Telefonica Research, Pl. Ernest Lluch, 5, 08019 Barcelona, Catalonia, Spain}
\affil[4]{Universitat de Barcelona, Diagonal, 690, 08034 Barcelona, Spain}
\affil[5]{BigML, Inc, 2851 NW 9th Suite, Conifer Plaza Building, Corvallis, OR 97330, US}

\bibliographystyle{apa}
\setcitestyle{authoryear,open={(},close={)}}
\maketitle

\begin{abstract}
Cash managers make daily decisions based on predicted monetary inflows from debtors and outflows to creditors. Usual assumptions on the statistical properties of daily net cash flow include normality, absence of correlation and stationarity. We provide a comprehensive study based on a real-world cash flow data set from small and medium companies, which is the most common type of companies in Europe. We also propose a new cross-validated test for time-series non-linearity showing that: (i) the usual assumption of normality, absence of correlation and stationarity hardly appear; (ii) non-linearity is often relevant for forecasting; and (iii) typical data transformations have little impact on linearity and normality. Our results provide a forecasting strategy for cash flow management which performs better than classical methods. This evidence may lead to consider a more data-driven approach such as time-series forecasting in an attempt to provide cash managers with expert systems in cash management.
\end{abstract}


\section{Introduction}

\begin{figure}[htb]
\centering
\includegraphics[width=0.9\textwidth]{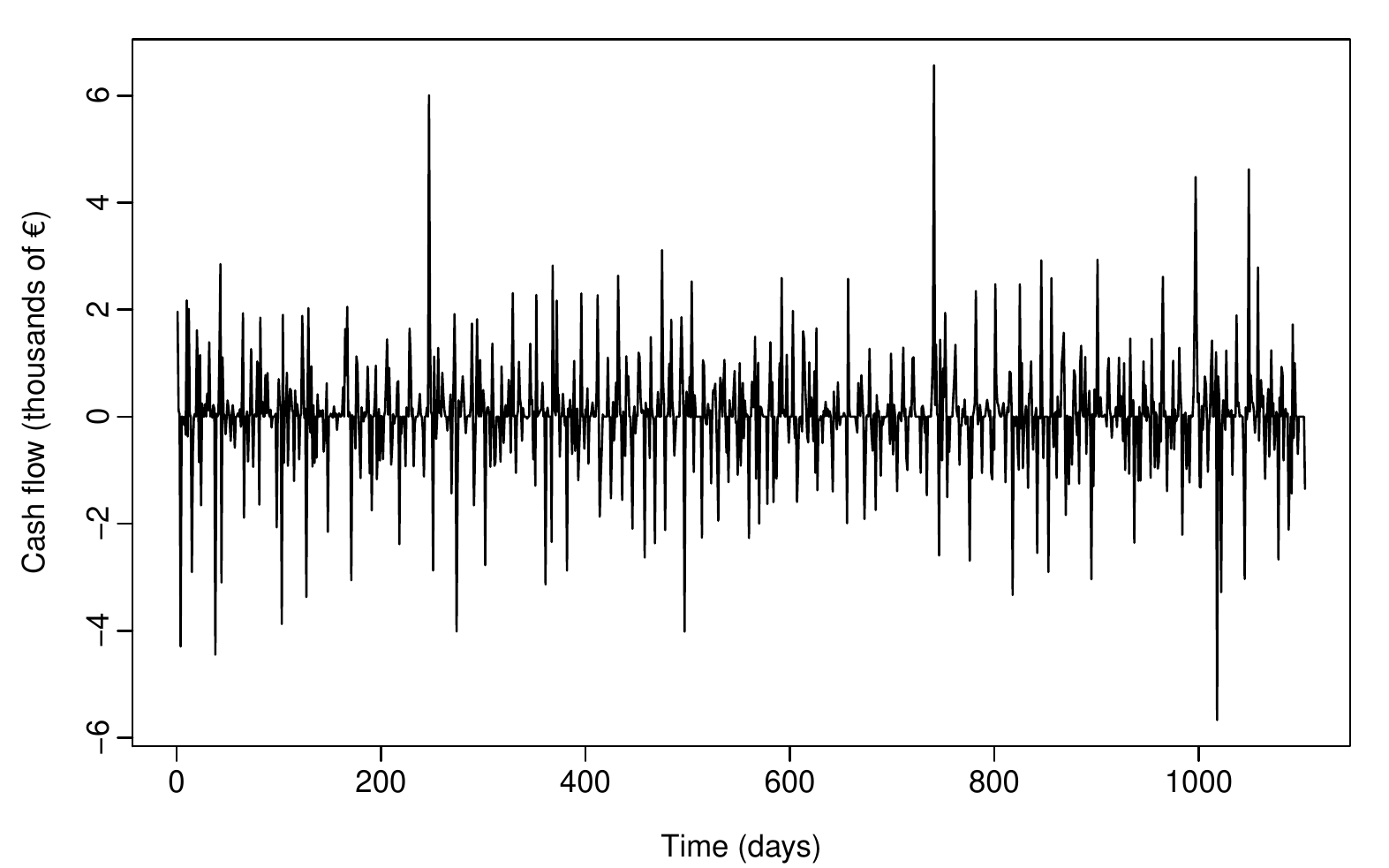}
\caption{\label{fig:example}Example of a cash flow time-series.}
\end{figure}

Cash management is concerned with the efficient use of a company's cash and short-term investments such as marketable securities. The focus is placed on maintaining the amount of available cash as low as possible, while still keeping the company operating efficiently. In addition, companies may place idle cash in short-term investments \citep{ross2002fundamentals}. Then, the cash management problem can be viewed as a trade-off between holding and transaction costs. If a company tries to keep balances too low, holding cost will be reduced, but undesirable situations of shortage will force to sell available marketable securities, hence increasing transaction costs. In contrast, if the balance is too high, low trading costs will be produced due to unexpected cash flow, but the company will carry high holding costs because no interest is earned on cash. Therefore, there is a target cash balance which each company must optimize according to the particular characteristics of its cash flows. An example of a cash flow time-series is shown in Figure~\ref{fig:example}. 

Since \cite{baumol1952transactions}, a number of cash management models have been proposed to control cash balances. These models are based either on the specific statistical properties of cash balances or on cash flow forecasts. A comprehensive review of models, from the first proposals to the most recent contributions, can be found in \cite{gregory1976cash,srinivasan1986deterministic}, and \cite{da2015stochastic}. Most of them are based on assuming a given probability distribution for cash flows such as: (i) a random walk in the form of independent Bernouilli trials as in \cite{miller1966model}; (ii) a Wiener process as in \cite{constantinides1978existence,premachandra2004diffusion}, and \cite{baccarin2009optimal}; (iii) a double exponential distribution as in \cite{penttinen1991myopic}. From these and other works, we observe that common assumptions on the statistical properties of cash flow time-series include:
\begin{itemize}
\item Normality: cash flows follow a Gaussian distribution with observations symmetrically centered around the mean, and with finite variance.
\item Absence of correlation: the occurrence of past cash flows does not affect the probability of occurrence of the next ones.
\item Stationarity: the probability distribution of cash flows does not change over time and, consequently, its statistical properties such as the mean and variance remain stable. 
\item Linearity: cash flows are proportional either to another (external) explanatory variable or to a combination of (external) explanatory variables.
\end{itemize}

Surprisingly, little and/or contradictory empirical evidence on these assumptions has been provided besides individual cases through time. Early on, negative normality tests were reported in \cite{homonoff1975cash} for the times series samples of a manufacturing company. Contrastingly, later on, \cite{emery1981some} reported normally distributed cash flow, after data transformation, for two out of three companies, and a small serial dependence for all of them. \cite{pindado1996evidencia} provided negative normality and independence results on 36 companies, but considering daily cash flow for only a single month. Previous works also reported day-of-week and day-of-month effects on cash flows, in line with the works of \cite{stone1977daily,miller1985daily}, and \cite{stone1987daily}. Recently, \cite{gormley2007utility} described the time-series from a multinational company with a non-normal distribution and serial dependence.

We consider that the evidence derived from these works is inconclusive due to: (i) the disagreement between the conclusions of some of the works; (ii) the limited number of companies analyzed; and (iii) the short time range of the observations. Moreover, none of the previous works considered the presence of non-linear patterns for forecasting purposes. In this work, we provide an analysis of the statistical properties of 54 real cash flow data sets from small and medium companies in Spain as a representative sample of the most common type of companies in Europe. Indeed, small and medium companies contribute to 99.8\% of all enterprises, 57.4\% of value added, and 66.8\% of employment across the EU28 \citep{muller2015annual}. To the best of our knowledge, this is the most comprehensive empirical study on daily cash flow so far. We base this statement on both the length and number of data sets, which amounts to 58005 observations in total, with a minimum, average and maximum time range of 170, 737, 1508 working days, respectively. In addition, we consider a wider range of statistical properties. A further contribution of the present work is to make all the aforementioned data publicly available online\footnote{\url{http://www.iiia.csic.es/~jar/54datasets3.csv}}. Finally, from a forecasting perspective, we also aim to identify the family of forecasters that best accommodate to cash flow time-series data sets. To this end, we propose a new and simple cross-validated test for non-linearity that provides further knowledge to cash managers in their search for better forecasting models. 

Our results show the unlikely occurrence of normality, absence of correlation and stationarity in the data sets under study. These results are consistent with the cited reports of \cite{homonoff1975cash}, based on only one time-series, and \cite{pindado1996evidencia}, based on a very short time range, raising doubts about the claim of independence. We also report that normality could not be achieved through removing outliers, contrary to what was reported by \cite{emery1981some}, based on only three time-series. Our analysis also confirms the influence of seasonality as suggested in \cite{miller1985daily} and \cite{stone1987daily}. Thus, we consider that our results provide stronger evidence against normality, uncorrelatedness and stationarity than previous works. Note that we do not claim that these results can be extrapolated to all kind of companies. On the contrary, we provide further evidence against standard assumptions in cash management. This evidence may lead to consider a more data-driven approach such as time-series forecasting in order to provide cash managers with expert systems in cash management \citep{nedovic2002expert}.

In an attempt to achieve Gaussian and stationary time-seres, practitioners typically use the Box-Cox transformation \citep{box1964analysis}, and time-series differencing \citep{makridakis2008forecasting}. Furthermore, some kind of outlier treatment is also a recommended practice. Then, we also study the impact of outlier treatment by replacing them with linear interpolations between two consecutive observations. However, in our study, we find little benefit when these methods are applied to our data sets. As a result, we point out the underlying question about data transformation in relation to the properties of a time-series. Is it always possible to achieve a Gaussian and linear time-series through data transformations? We here rely both on common statistical tests and on our novel non-linearity test to answer this question and we find that: (i) outlier treatment and Box-Cox transformation are not always enough to achieve normality; (ii) outlier treatment produces mixed results in terms of noise reduction and information loss; (iii) outlier treatment and Box-Cox transformations do not produce linearity. These results suggest that non-linear models conform a justifiable alternative for cash flow time-series forecasting, beyond the current conjectures of the literature.

The remaining of the paper is organized as follows. In Section \ref{sec:stat}, we provide a statistical summary of the contributed 54 real cash flow data sets including normality, independence and stationarity. In Section \ref{sec:linearity}, we propose a new cross-validated test for non-linearity based on the comparison of a linear model and a non-linear model. Later, we present in Section \ref{sec:outliers} detailed results on the impact of data transformations on linearity. Finally, we provide some concluding remarks in Section \ref{sec:conclusions}.

\section{Data summary \label{sec:stat}}

The data set contains daily cash flows from 54 different companies from the manufacturing and the service sector in Spain with annual revenue up to \euro10 million each. No company from the primary sectors is included in the sample. We select only small and medium companies since it is the most common size of companies in both Spain and Europe \citep{muller2015annual}. This data set covers a date range of about 8 years and is available online. An instance in the data set contains the following fields or columns:

\begin{itemize}
\item Date: standardized YYYY-MM-DD dates from 2009-01-01 to 2016-28-08. 
\item Company: company identifier from 1 to 54.
\item NetCF: daily net cash flow in thousands of \euro.
\item DayMonth: categorical variable with the day of the month from 1 to 31.
\item DayWeek: categorical variable with the day of the week from 1 (Monday) to 7 (Sunday).
\end{itemize}

Table \ref{tab:sum} shows the statistical summary of daily net cash flow on non-holidays, grouped by company. Small and medium companies are likely to experiment daily null cash flows, meaning that no monetary movement is observed at a particular working day even under regular activity. As a result, the occurrence of null cash flows is an important characteristic of small and medium companies due to the size of companies. Indeed, almost 30\% of the companies in our data set present more than 25\% of null cash flow observations even at working days. This fact implies that a null cash flow prediction will be right at least 25\% of the times for this group of data sets. Therefore, two good baseline forecasting models for comparative purposes would be an \textit{always-predict-null} or an \textit{always-predict-mean} forecaster \citep{makridakis2008forecasting}. 

In addition, the average net cash flow shows that a high percentage of companies present either positive or negative drift with the exception of companies 5 and 28. High positive kurtosis indicates a peaked data distribution in comparison to the normal distribution that has zero kurtosis. The skewness is a measure of the symmetry of the data distribution. Negative skewness indicates that the left tail is longer, and positive skewness indicates that the right tail is longer.

\begin{table}[htp]
  \tiny
  \centering
  \caption{\label{tab:sum}Data sets statistical summary. Mean, standard deviation, minimum, maximum in thousands of \euro.}
  \begin{tabular}{lrrrrrrrr}
    Id & Length & Null \%  & Mean  & Std   & Kurtosis  & Skewness  & Min   & Max \\
    \hline
    1     & 856   & 35,7  & 0,01  & 3,38  & 594,81 & 22,37 & -9,07 & 90,27 \\
    2     & 684   & 29,8  & 0,26  & 5,80  & 58,98 & 3,69  & -56,51 & 62,66 \\
    3     & 856   & 8,5   & 0,36  & 35,35 & 163,62 & 6,28  & -303,20 & 671,04 \\
    4     & 1201  & 34,9  & -0,12 & 14,32 & 78,14 & -6,30 & -223,38 & 72,76 \\
    5     & 849   & 19,4  & 0,00  & 1,67  & 56,10 & -0,48 & -18,26 & 16,42 \\
    6     & 799   & 20,7  & 0,01  & 6,63  & 33,21 & -2,42 & -68,97 & 56,27 \\
    7     & 772   & 38,5  & 0,07  & 5,36  & 86,75 & 6,74  & -24,41 & 82,91 \\
    8     & 695   & 21,7  & 0,05  & 3,15  & 14,27 & -2,57 & -24,21 & 11,31 \\
    9     & 852   & 18,8  & 0,73  & 56,54 & 18,92 & -0,78 & -411,41 & 473,36 \\
    10    & 744   & 13,2  & 0,12  & 6,95  & 70,63 & 0,60  & -81,13 & 78,72 \\
    11    & 639   & 62,6  & -0,05 & 8,56  & 391,86 & -17,65 & -191,53 & 30,74 \\
    12    & 503   & 2,6   & 0,48  & 35,30 & 449,38 & 20,70 & -47,27 & 771,38 \\
    13    & 697   & 24,7  & 0,52  & 24,24 & 18,81 & 2,06  & -99,39 & 227,45 \\
    14    & 604   & 4,6   & 0,10  & 13,23 & 8,51  & 1,05  & -63,23 & 92,71 \\
    15    & 605   & 4,1   & 0,68  & 11,67 & 4,43  & 0,33  & -54,75 & 55,61 \\
    16    & 596   & 6,4   & 0,01  & 1,46  & 107,82 & 6,68  & -8,48 & 22,61 \\
    17    & 1102  & 25,1  & 0,58  & 13,31 & 215,97 & 11,96 & -118,01 & 250,13 \\
    18    & 552   & 3,1   & 0,16  & 2,16  & 70,23 & 5,10  & -16,14 & 26,36 \\
    19    & 503   & 2,4   & -0,31 & 2,58  & 6,43  & 0,50  & -15,06 & 15,28 \\
    20    & 848   & 27,8  & 0,02  & 1,07  & 96,19 & 3,86  & -12,07 & 16,04 \\
    21    & 829   & 18,7  & -0,06 & 5,99  & 33,36 & -1,62 & -70,00 & 53,17 \\
    22    & 494   & 1,6   & -0,46 & 27,28 & 22,64 & -1,96 & -244,29 & 138,87 \\
    23    & 604   & 9,1   & 1,63  & 20,85 & 79,99 & 5,41  & -124,19 & 269,27 \\
    24    & 1097  & 8,4   & 0,96  & 20,36 & 95,45 & 6,48  & -73,33 & 317,85 \\
    25    & 587   & 10,9  & 0,49  & 13,94 & 119,60 & 6,93  & -116,01 & 201,13 \\
    26    & 751   & 11,6  & -0,02 & 1,77  & 15,73 & 0,15  & -10,73 & 15,56 \\
    27    & 332   & 8,1   & 0,29  & 1,64  & 10,60 & 2,14  & -4,36 & 11,84 \\
    28    & 855   & 5,1   & 0,00  & 4,64  & 13,83 & 1,77  & -18,10 & 39,01 \\
    29    & 609   & 13,6  & 0,04  & 6,07  & 108,66 & -6,35 & -90,04 & 55,89 \\
    30    & 554   & 8,1   & 0,03  & 1,47  & 68,26 & 5,47  & -4,81 & 19,82 \\
    31    & 372   & 29,6  & 0,37  & 8,05  & 31,46 & -2,41 & -80,44 & 34,95 \\
    32    & 1103  & 24,8  & 0,28  & 4,03  & 11,07 & 0,54  & -25,76 & 24,50 \\
    33    & 854   & 31,0  & -0,19 & 6,81  & 115,63 & -1,74 & -94,33 & 95,59 \\
    34    & 1508  & 11,5  & -0,06 & 10,13 & 19,89 & -2,32 & -96,82 & 49,65 \\
    35    & 501   & 7,4   & 0,20  & 5,40  & 11,41 & -0,58 & -31,42 & 29,19 \\
    36    & 359   & 11,4  & 0,42  & 1,85  & 12,24 & 2,44  & -7,87 & 11,84 \\
    37    & 361   & 3,0   & -0,69 & 17,82 & 139,06 & -1,38 & -228,88 & 218,42 \\
    38    & 170   & 9,4   & -1,20 & 7,10  & 43,34 & -5,73 & -61,93 & 19,66 \\
    39    & 1104  & 29,0  & 0,02  & 0,95  & 7,95  & -0,07 & -5,67 & 6,57 \\
    40    & 198   & 0,0   & 0,78  & 12,38 & 0,58  & 1,02  & -25,63 & 36,91 \\
    41    & 341   & 17,6  & -0,25 & 8,34  & 15,80 & 1,22  & -44,29 & 64,34 \\
    42    & 566   & 11,0  & 0,01  & 1,82  & 308,62 & -15,80 & -37,02 & 7,48 \\
    43    & 750   & 3,2   & 0,34  & 13,10 & 7,66  & -0,04 & -65,84 & 73,40 \\
    44    & 287   & 4,2   & 0,52  & 11,46 & 81,19 & -0,05 & -118,74 & 120,34 \\
    45    & 1465  & 49,8  & 0,04  & 9,12  & 43,51 & -2,89 & -107,20 & 75,47 \\
    46    & 565   & 44,8  & 0,54  & 5,58  & 75,41 & 2,91  & -51,16 & 73,83 \\
    47    & 503   & 4,4   & 1,98  & 46,81 & 46,03 & 1,37  & -338,39 & 478,26 \\
    48    & 605   & 13,1  & 0,21  & 22,71 & 34,31 & -1,68 & -207,04 & 203,09 \\
    49    & 993   & 50,5  & -0,08 & 1,36  & 27,18 & -2,18 & -10,78 & 12,73 \\
    50    & 605   & 45,0  & -0,01 & 27,37 & 43,79 & -2,01 & -262,52 & 221,96 \\
    51    & 1225  & 0,2   & 15,09 & 96,96 & 2,77  & 0,12  & -419,88 & 481,66 \\
    52    & 1225  & 0,4   & 8,94  & 49,39 & 36,23 & 2,81  & -325,46 & 700,66 \\
    53    & 1223  & 39,7  & 0,47  & 9,13  & 203,12 & -10,25 & -196,88 & 38,48 \\
    54    & 1225  & 52,3  & 0,46  & 77,91 & 151,93 & 4,28  & -1021,36 & 1532,10 \\
    \hline
   \end{tabular}
\end{table}%

\subsection{Normality}

First, we study if our cash flows follow a Gaussian distribution. In fact, the observed kurtosis and skewness can be used as a first normality test of the data distribution for each company. Table \ref{tab:sum} shows that no company presents zero kurtosis and skewness. Only company 40, with kurtosis 0.58 and skewness 1.02, could be considered close to normality. Two additional tests can be used to either verify or reject the hypothesis of normality: the Shapiro-Wilk test for normality \citep{royston1982extension} and the Lilliefors (Kolmogorov-Smirnov) test for normality \citep{lilliefors1967kolmogorov}. The results from these two tests allow us to reject the hypothesis of normally distributed cash flows for all the companies in our data set (no exception). 

As pointed out elsewhere \citep{emery1981some,pindado1996evidencia}, a possible explanation for non-normality could be the presence of abnormally high values or heavy tails. Thus, we repeated the Shapiro-Wilk test and the Lilliefors (Kolmogorov-Simirnov) test for normality, but using a trimmed version of the net cash flow time-series by deleting observations greater or lower than three times the sample standard deviation. No difference in the results of the tests was observed, confirming the non-normality hypothesis beyond the conjectures of \cite{emery1981some} and \cite{pindado1996evidencia}.

Non-normal residuals may be problematic in the estimation process when using linear models. Data transformations such as the \cite{box1964analysis} transformation to normality represent a possible solution. Forecasts are then calculated on the transformed data, but we must reverse the transformation to obtain forecasts on the original data, resulting in two additional steps. However, these transformations are not always the solution to the non-normality problem. Using both the original observations and the trimmed version of our data sets, we proceeded to transform the data using a Box-Cox transformation of the type:
\begin{equation}
y^{(\lambda)}=\left\{\begin{array}{lll} \frac{(y+ \lambda_2)^{\lambda_1} -1}{\lambda_1}  & \mbox{if} & \lambda_1 \neq 0, \\ \operatorname{log}(y+\lambda_2) & \mbox{if} & \lambda_1 = 0,  \end{array}\right.
\label{eq:boxcox}
\end{equation}
where $y$ is the original time-series, and $\lambda_1$ and $\lambda_2$ are parameters. In these experiments, we first set $\lambda_2$ to minus two times the minimum value of the time-series to avoid problems with negative and zero observations. \cite{box1964analysis} provided the profile likelihood function for $\lambda_1$ and suggested to use this function as a way to tune this parameter. Then, we follow the recommendations in \cite{venables2013modern} to compute the profile likelihood function for $\lambda_1$, and we later select the value that maximizes the log-likelihood function when applying a linear regression model of the time-series based on day-of-month and day-of-week dummy variables. Then, we repeated the Lilliefors (Kolmogorov-Smirnov) test for normality obtaining again negative results for normality, both for the original and for the trimmed version of our data sets. As a result, we must conclude that, even after Box-Cox transformation, the normality hypothesis does not hold.

\subsection{Correlation and seasonality}

In what follows, we test the correlation of cash flows and we also explore if seasonality is present. Autoregressive Integrated Moving Average (ARIMA) models by \cite{box1976time}, have been extensively used for time-series analysis and forecasting. When dealing with time-series, the autocorrelation coefficient, $r_k$, describes the relationship between observations that are lagged $k$ time periods \citep{makridakis2008forecasting}. We say that a time-series is not autocorrelated when the $r_k$ values for different lags are close to zero. An example of an independent time-series is the so-called white-noise model where each observation is made by adding a random component to a certain level. 

\begin{figure}[htb]
\centering
\includegraphics[width=1\textwidth]{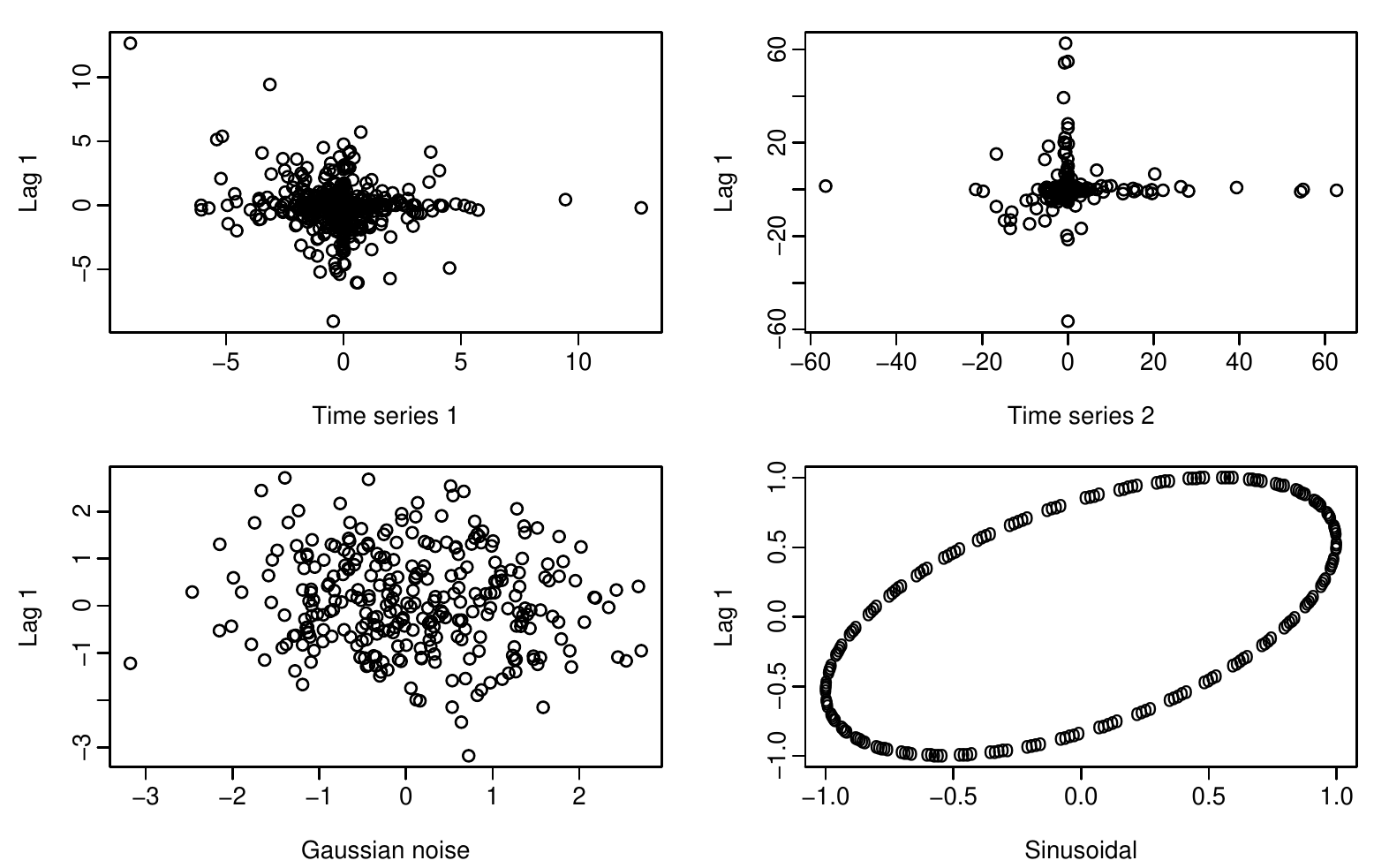}
\caption{\label{fig:delay}Poincaré map with lag 1 for time-series 1 and 2.}
\end{figure}

An intuitive plot to assess correlation is the Poincaré map \citep{kantz2004nonlinear}, which is a scatter plot of the original time-series and a $k$-periods lagged time-series as in Figure \ref{fig:delay}, which shows a lag of 1 day for time-series 1 and 2 from Table \ref{tab:sum}. As a reference, we also include the Poincaré map for a white-noise and for a sinusoidal time-series. A cloud of points suggests lack of correlation, as for time-series 1 and white-noise, and the presence of any form suggests a more complex relationship, as for time-series 2 and the sinusoidal.

A more general approach is to consider a set of the first $r_k$ values as a whole as in the \cite{ljung1978measure} test, which we applied to our data and produced mixed results. More precisely, we found that the null hypothesis of independence could not be rejected in 24 out of 54 companies as summarized in Table \ref{tab:indep}. These results imply that some kind of serial correlation is likely to be present in the case of companies presenting a certain degree of autocorrelation in the sample. A plausible type of serial correlation is seasonality, that is, the existence of a pattern that repeats itself over fixed time intervals in the data \citep{makridakis2008forecasting}. It can be identified by significant autocorrelation coefficients. Seasonal trend decomposition methods \citep{cleveland1990stl}, seasonal ARIMA models \citep{box1976time,franses2005forecasting} or linear (and non-linear) regression models based on seasonal variables are available options to deal with seasonality. In cash flow forecasting, the distribution approach by \cite{miller1985daily} also deserves to be mentioned.

As mentioned in the introduction, previous works by \cite{emery1981some,miller1985daily,stone1987daily}; and \cite{pindado1996evidencia}, reported the influence of day-of-month and day-of-week effects on cash flow patterns. Here, we test the presence of seasonality by fitting a regression model on daily cash flows using day-of-month and day-of-week dummy variables. Table \ref{tab:indep} reports, on the one hand, the Ljung-Box correlation test and, on the other hand, the F-statistic, the $p$-value and the coefficient of determination $R^2$, derived from the regression. Interestingly, not all the data sets whose Ljung-Box tests were rejected implied better regressions than the baseline based on dummy variables. Non-linear patterns, non-periodical temporal correlations, and the effect of outliers become possible explanations as we will see below.

\begin{table}[!htb]
  \tiny
  \centering
  \caption{\label{tab:indep}Correlation and seasonality test results.}
  \begin{tabular}{ccccc}
    Id & Ljung-Box test & F-statistic & p-value & $R^2$ \\
    \hline
    1     & Non-rejected & 1,99  & <0,05 & 0,08 \\
    2     & Rejected & 1,05  & 0,39  & 0,05 \\
    3     & Non-rejected & 1,87  & <0,05 & 0,07 \\
    4     & Rejected & 1,51  & <0,05 & 0,04 \\
    5     & Rejected & 1,85  & <0,05 & 0,07 \\
    6     & Non-rejected & 1,12  & 0,29  & 0,05 \\
    7     & Rejected & 5,47  & <0,05 & 0,20 \\
    8     & Rejected & 0,79  & 0,80  & 0,04 \\
    9     & Rejected & 5,30  & <0,05 & 0,18 \\
    10    & Rejected & 2,04  & <0,05 & 0,09 \\
    11    & Non-rejected & 0,97  & 0,51  & 0,05 \\
    12    & Non-rejected & 0,98  & 0,51  & 0,07 \\
    13    & Rejected & 5,21  & <0,05 & 0,21 \\
    14    & Rejected & 7,13  & <0,05 & 0,30 \\
    15    & Rejected & 1,92  & <0,05 & 0,10 \\
    16    & Non-rejected & 4,31  & <0,05 & 0,21 \\
    17    & Non-rejected & 4,91  & <0,05 & 0,14 \\
    18    & Rejected & 2,99  & <0,05 & 0,16 \\
    19    & Rejected & 2,58  & <0,05 & 0,16 \\
    20    & Non-rejected & 2,71  & <0,05 & 0,10 \\
    21    & Non-rejected & 1,37  & 0,08  & 0,06 \\
    22    & Non-rejected & 1,49  & <0,05 & 0,10 \\
    23    & Rejected & 5,60  & <0,05 & 0,25 \\
    24    & Non-rejected & 15,41 & <0,05 & 0,33 \\
    25    & Non-rejected & 4,23  & <0,05 & 0,21 \\
    26    & Rejected & 1,22  & 0,18  & 0,05 \\
    27    & Non-rejected & 1,24  & 0,18  & 0,12 \\
    28    & Rejected & 5,64  & <0,05 & 0,19 \\
    29    & Non-rejected & 1,37  & 0,08  & 0,08 \\
    30    & Rejected & 6,18  & <0,05 & 0,29 \\
    31    & Non-rejected & 1,25  & 0,16  & 0,11 \\
    32    & Rejected & 4,81  & <0,05 & 0,13 \\
    33    & Rejected & 1,57  & <0,05 & 0,06 \\
    34    & Rejected & 11,61 & <0,05 & 0,21 \\
    35    & Rejected & 0,99  & 0,49  & 0,07 \\
    36    & Non-rejected & 1,82  & <0,05 & 0,16 \\
    37    & Rejected & 1,58  & <0,05 & 0,14 \\
    38    & Non-rejected & 1,06  & 0,39  & 0,21 \\
    39    & Rejected & 6,11  & <0,05 & 0,16 \\
    40    & Rejected & 0,86  & 0,68  & 0,15 \\
    41    & Non-rejected & 1,72  & <0,05 & 0,16 \\
    42    & Non-rejected & 3,90  & <0,05 & 0,20 \\
    43    & Rejected & 2,96  & <0,05 & 0,12 \\
    44    & Non-rejected & 1,89  & <0,05 & 0,20 \\
    45    & Rejected & 1,26  & 0,15  & 0,03 \\
    46    & Non-rejected & 1,32  & 0,11  & 0,08 \\
    47    & Non-rejected & 0,90  & 0,63  & 0,06 \\
    48    & Non-rejected & 1,71  & <0,05 & 0,09 \\
    49    & Rejected & 26,15 & <0,05 & 0,48 \\
    50    & Rejected & 1,24  & 0,17  & 0,07 \\
    51    & Rejected & 16,66 & <0,05 & 0,32 \\
    52    & Rejected & 5,01  & <0,05 & 0,13 \\
    53    & Non-rejected & 1,59  & <0,05 & 0,04 \\
    54    & Rejected & 0,88  & 0,67  & 0,02 \\
    \hline
    \end{tabular}
\end{table}%

\subsection{Stationarity \label{sec:stationarity}}

In this section, we analyze if cash flows from our data set can be labeled as stationary. Basically, stationarity means that there is no drift in the time-series behavior over time. We can visually assess stationarity by inspecting a time-series plot as the one shown in Figure \ref{fig:example}. Virtually, every process we find in nature is non-stationary, since its parameters depend on time \citep{kantz2004nonlinear}. However, a minimum requirement is that basic statistical properties of a distribution, such as mean and variance, remain constant over time, when measured through appropriately long time windows. 

\cite{emery1981some} studied stationarity by applying the Kolmogorov-Smirnov test for normality of cash flow by months. For comparative purposes, we applied the same procedure and, if any of the monthly tests rejected the hypothesis of normality, the whole time-series was considered non-stationary. Following this procedure, only company 43 could be considered stationary.

Following the recommendations in \cite{kantz2004nonlinear}, we also perform a stationarity test based on the fluctuations of a sample mean and variance. More precisely, we compute the sample mean and variance of each time-series by months and obtain the standard errors for both. If the observed fluctuations of the running mean and variance are within these errors, then we consider the time-series stationary. The results from this test shows that none of the time-series is stationary.

One way of removing non-stationarity is time-series differencing, which can be defined as the change between two consecutive observations. After differencing, we repeated our simple test obtaining slightly different results but none of them can be considered stationary. Thus, as a result, we conclude that our cash flow time-series are non-stationary, even after differencing.

\subsection{Discussion}

Our results show that the widely extended hypothesis of cash flow normality is not present in our data sets. The presence of high abnormal values does not explain this behavior since non-normality persisted after removing these abnormal values. Non-linearity could be a possible explanation as we will see below. We also reported mixed results on autocorrelation and the influence of day-of-month and day-of-week effects on cash flow along the lines of the literature. We additionally report that common solutions to non-normality and non-stationarity such as data transformation and differencing produced little benefit when applied to our time-series. Since seasonality and serial correlation are also present in our data set, we further explore the usefulness of alternative forecasting models. More precisely, we next study linearity and data transformation as an additional part of our empirical analysis for cash flow forecasting.

\section{A simple cross-validated test for non-linearity \label{sec:linearity}}

Most forecasting models are linear for computational convenience. However, non-linear patterns are likely to be present in finance and business time-series. A time-series linear model is defined as a variable $y_t$ that depends on the additive contribution of a number of explanatory variables in vector $\boldsymbol{x_t}$ for any time $t$ as follows:
\begin{equation}
y_t = \boldsymbol{\beta^{\prime}} \boldsymbol{x_t}+e_t
\end{equation}
where $\boldsymbol{\beta^\prime}$ is a transposed vector of coefficients, and $e_t$ is the error or the residual component. An alternative and more general model can also be considered:
\begin{equation}
y_t = g(\boldsymbol{x_t}) + \epsilon_t
\label{eq:non-linear}
\end{equation}
where $g(\boldsymbol{x_t})$ is any function that aims to describe the underlying time-series. By considering non-linear relationships between the set of predictors and the cash flow dependent variable, more complex patterns such as interactions between the day-of-week and the day-of-month may be captured.

Different tests of linearity can be found in \cite{ramsey1969tests,keenan1985tukey,lee1993testing}, and \cite{castle2010low}. Basically, all of them follow a common approach: first, they choose a function $g(\boldsymbol{x_t})$ in equation (\ref{eq:non-linear}) including linear and non-linear terms and, second, they test for the significance of the non-linear terms. However, these approaches do not accommodate well for forecasting purposes due to the following reasons: (i) the assumption of a specific form $g(\boldsymbol{x_t})$ for the regression equation such as quadratic, cubic or exponential forms; (ii) cross-validation is neglected. 

\begin{figure}[!htb]
\centering
\begin{tikzpicture}[node distance = 3cm, auto]
    \node [print, text width=2cm, node distance=5.5cm] (ts) {time-series};
    \node [block, below of=ts, node distance= 2.5cm] (cv) {Cross-validation Algorithm 1};
    \node [decision, below of=cv, node distance =3.5cm, text width=3cm] (trivial) {Better than baseline?};
    \node [decision, below of=trivial, node distance= 3.5cm, text width=2.8cm] (wilcoxon) {Better than linear?};
	\node [print, left of=ts, node distance= 3.5cm] (linear) {Linear model};
    \node [print, right of=ts, node distance= 3.5cm, text width=2.5cm] (non-linear) {Non-linear model};
    \node [block, below of=non-linear, node distance=11cm, text width=2.5cm] (labelasnl) {Label as non-linear};
    \node [block, below of=linear, node distance=11cm, text width=2.5cm] (labelasl) {Label as linear};
    \node [block, below of=non-linear, node distance=7.5cm, text width=2.5cm] (labelast) {Label as trivial};
  
   \path [line] (ts) -- node [anchor=east] {} (cv);
   \path [line] (cv) -- node [anchor=west, text width=3cm] {one error sample for each model} (trivial);
   \path [line] (trivial) -- node [anchor=west] {Yes} (wilcoxon);
   \path [line] (non-linear) |- node [near end, anchor=east] {} (cv);
   \path [line] (linear) |- node [near end, anchor=east] {} (cv);
   \path [line] (wilcoxon) -| node [near end, anchor=west] {Yes} (labelasnl);
   \path [line] (wilcoxon) -| node [near end, anchor=west] {No} (labelasl);
   \path [line] (trivial) -| node [near end, anchor=west] {No} (labelast);
\end{tikzpicture}
\caption{\label{fig:flow} Simplified flow chart for our cross-validated test for non-linearity.}
\end{figure}
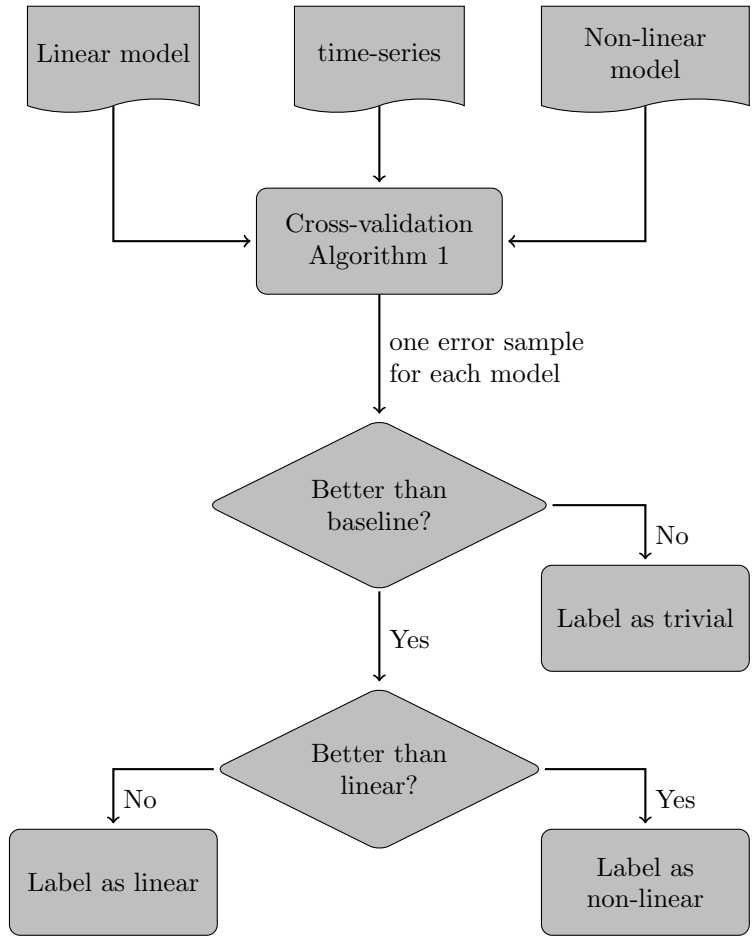

If we relax the assumption of linearity, different non-linear models such as random forests \citep{breiman2001random}, neural networks \citep{hornik1989multilayer,zhang1998forecasting}, or radial basis functions \citep{broomhead1988multivariable}, could also be considered. However, the consideration of non-linear functions may lead to overfitting to the original time-series. To prevent this problem, we propose the use of time-series cross-validation. Cross-validation is a method to assess the predictive performance of a forecasting model that circumvents the problem of overfitting the data by testing the accuracy of the model on subset of data not used in the estimation \citep{hyndman2013forecasting}. As a result, we here propose a simple cross-validated test for non-linearity based on the following steps:
\begin{enumerate}
\item Estimate two alternative forecasting models, one linear and another one non-linear.
\item Cross-validate the predictive accuracy of both models with respect to a baseline.
\item Label as trivial if both models are significantly worse than the baseline.
\item Label as non-linear if the error of the non-linear model is significantly lower than that of the linear model. Otherwise, label as linear as described in Figure \ref{fig:flow}.
\end{enumerate}

Since we do not assume any distribution for the forecasting results, we use the two-sided Wilcoxon rank-based for statistically significant differences in performance between models. More precisely, we test the null hypothesis that the distribution of the difference is symmetric about zero with a 95\% confidence interval \citep{wilcoxon1970critical}. Approximate $p$-values are computed based on the asymptotic distribution of the two-sided Wilcoxon test statistic and used to label data sets as detailed in Algorithm \ref{GeneralAlg}. 

\begin{algorithm}
\caption{Algorithm for a simple cross-validation test for non-linearity \label{GeneralAlg}}
  \begin{algorithmic}[1]
    \State \textbf{Input:} Cash flow data set of \emph{T} instances, minimum number \emph{k} of instances to estimate a model, baseline $m_0$, linear model $m_1$, non-linear model $m_2$, prediction horizon~\emph{h}, level of significance $\alpha$.  
    \State \textbf{Output:} Average prediction error, statistic for the difference in mean errors, confidence interval.
    \For{$i=1,2,\ldots,T-k-h+1$}
    	\State   Select the instances from time $k+i$ to $k+h+i-1$, for the test set;
  		\State   Estimate $m_0$ with instances at times $1,2,\ldots,k+i-1$;
  		\State   Estimate $m_1$ with instances at times $1,2,\ldots,k+i-1$;
  		\State   Estimate $m_2$ with instances at times $1,2,\ldots,k+i-1$;
  		\State   Compute test errors $\varepsilon_0$, $\varepsilon_1$, $\varepsilon_2$ from time $k+i$ to $k+h+i-1$;
    \EndFor
    \State Compute average \emph{h}-step errors $\varepsilon_0(h)$, $\varepsilon_1(h)$, $\varepsilon_2(h)$;
    \State Test for $\alpha$ significant differences between $\varepsilon_0(h)$, $\varepsilon_1(h)$, $\varepsilon_2(h)$;
    \If{$\varepsilon_0(h) < \varepsilon_1(h)$ \textbf{and} $\varepsilon_0(h) < \varepsilon_2(h)$}
    	\State Label as trivial;
    \ElsIf{$\varepsilon_2(h) < \varepsilon_1(h)$} 
    	\State Label as non-linear;
    \Else
    	\State Label as linear.
    \EndIf
  \end{algorithmic}
\end{algorithm}


A common practice to assess the usefulness of forecasts derived from any model is to compare its accuracy to that of a baseline forecasting model. Thee use of a baseline model allows us to label our data sets as trivial if neither the linear model nor the non-linear model are able to improve the accuracy of the baseline. We here report accuracy results with respect to a mean forecaster, meaning that forecasts are always the average of all past observations. We also tried with an additional baseline forecaster using the last observed value as a forecast (persistence model) with much worse results in comparison to the mean forecaster.

We consider the minimum length $k$ to estimate a model as the 80\% of the oldest instances forming the training set. The remaining 20\% of the instances form the test set for cross-validation. Initially, both the linear and the non-linear model are estimated using the first 80\% of the instances. Then, forecasts for a prediction horizon up to 20 days are computed using the estimated models and squared errors are recorded. Later, we repeat the process by considering an increasing window with an additional instance in the training set to estimate the models. Again, forecasting errors are recorded for each remaining observation in the test set resulting into two paired error samples, one for the linear model and one for non-linear model.

A critical point when using our cross-validated test for non-linearity is the selection of both the linear and the non-linear forecast model. In essence, our test is a comparative tool based on forecasting accuracy as a proxy for non-linearity. Given a set of explanatory variables, a linear label result from our test implies that the non-linear model is not able to capture non-linearity. However, chances are that alternative non-linear models might perform differently. In this sense, if the time-series is not a white-noise process, then the search for a more informative set of features is meant to play a key role. As a result, multiple runs of our test are necessary to discard/assess non-linearity by using alternative linear and non-linear models

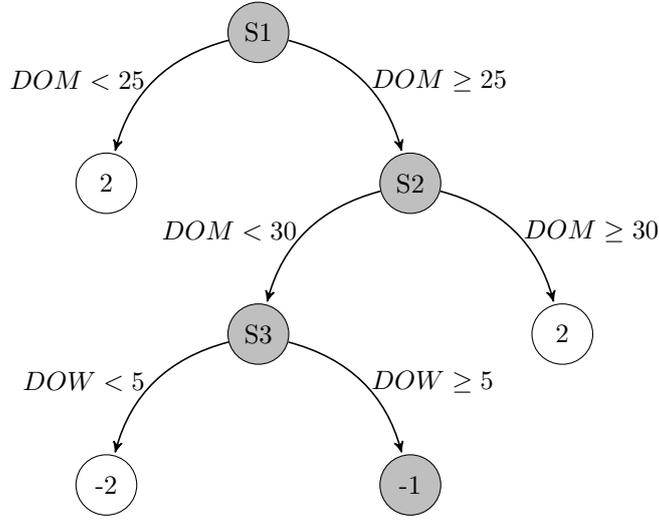
\begin{figure}[!htb]
\centering
\begin{tikzpicture}[node distance = 1cm]
\node[place] (level0) [fill=gray!50, minimum size=8mm]  {S1};
\node[place] (level1right) [below of = level0, right of= level0, fill=gray!50, minimum size=8mm]  {S2}
	edge[pre, bend right] node[right] {$DOM \geq 25$}  (level0);
\node[place] (level1left) [below of = level0, left of= level0, minimum size=8mm]  {2}
	edge[pre, bend left] node[left] {$DOM<25$}  (level0);
\node[place] (level2right) [below of = level1right, right of= level1right, minimum size=8mm]  {2}
	edge[pre, bend right] node[right] {$DOM \geq 30$}  (level1right);
\node[place] (level2left) [below of = level1right, left of= level1right, fill=gray!50, minimum size=8mm]  {S3}
	edge[pre, bend left] node[left] {$DOM < 30$}  (level1right);
\node[place] (level3right) [below of = level2left, right of= level2left, fill=gray!50, minimum size=8mm]  {-1}
	edge[pre, bend right] node[right] {$DOW \geq 5$}  (level2left);
\node[place] (level3left) [below of = level2left, left of= level2left, minimum size=8mm]  {-2}
	edge[pre, bend left] node[left] {$DOW < 5$}  (level2left);
\end{tikzpicture}
\caption{\label{fig:tree}A basic decision tree. DOM = Day-of-month; DOW = Day-of-week}
\end{figure}

For illustrative purposes, we here restrict ourselves to a linear regression model and a non-linear random forest model, both using day-of-month and day-of-week variables as predictors. In the case of the linear regression model, each instance contains 34 dummy predictor variables, 30 for day-of-month and 4 for day-of week, and a cash flow observation. In the case of random forests, each instance contains 2 categorical variables, one for day-of-month and one for day-of-week. Random forests are ensembles of slightly different decision trees \citep{ho1998random,breiman2001random}. An ensemble methodology is able to construct a predictive model by integrating multiple trees in what is called a decision forest \citep{dietterich2000ensemble}. Decision trees split the input space in subsets based on the value of features such as the day-of-month and day-of-month. In the example in Figure~\ref{fig:tree}, for days comprised between the 25th (node S1) and the 29th of each month (node S2) occurring on Friday (node S3), the predicted cash flow is -1.

Recent examples of time-series forecasting using random forests can be found in \cite{booth2014automated,zagorecki2015prediction} and \cite{salas2017empowering}. Summarizing, random forests are used to forecast variables based on an ensemble of different trees. Unlike linear regression, random forests allow to capture (if any) more complex relationships between predictor variables allowing us to identify possible non-linearities in the underlying cash flow process represented by our sample data sets.

Our results, summarized in Table \ref{tab:linearity}, show that only about half of the data sets can be labeled as trivial because neither the linear model nor the non-linear model were able to significantly beat the trivial forecaster. From those time-series in which the absence of correlation could no be rejected (see Ljung-Box test Table \ref{tab:indep}), 20 out of 24 were labeled as trivial. On the other hand, only 6 of them were labeled as non-linear according to our cross-validated definition. As mentioned above, these results depend on the selected forecasting models. Instead of claiming that random forests are able to better capture non-linear patterns than alternative models, we encourage practitioners to consider additional combinations of both linear and non-linear models.


\begin{table}[!htb]
  \centering
  \tiny
  \caption{\label{tab:linearity}Results of the test for non-linearity. Reg NSE = Regression normalized squared error; RF NSE = Random Forest normalized squared error }
    \begin{tabular}{crrrccc}
    \hline
    Id & Reg NSE   & RF NSE  & Statistic & p-value & Triviality & Linearity \\
    \hline
    1     & 0,99  & 1,00  & 26    & <0,05 & Non-Trivial & Linear \\
    3     & 0,99  & 1,01  & 8     & <0,05 & Non-Trivial & Linear \\
    4     & 1,00  & 1,01  & 0     & <0,05 & Non-Trivial & Linear \\
    7     & 0,81  & 0,83  & 0     & <0,05 & Non-Trivial & Linear \\
    9     & 0,90  & 0,93  & 3     & <0,05 & Non-Trivial & Linear \\
    13    & 0,86  & 0,88  & 13    & <0,05 & Non-Trivial & Linear \\
    14    & 0,76  & 0,77  & 45    & <0,05 & Non-Trivial & Linear \\
    16    & 0,85  & 0,86  & 64    & 0,13  & Non-Trivial & Linear \\
    18    & 0,86  & 0,88  & 63    & 0,12  & Non-Trivial & Linear \\
    19    & 0,96  & 0,94  & 182   & <0,05 & Non-Trivial & Non-linear \\
    20    & 0,99  & 0,98  & 209   & <0,05 & Non-Trivial & Non-linear \\
    23    & 0,78  & 0,79  & 78    & 0,33  & Non-Trivial & Linear \\
    24    & 0,73  & 0,79  & 0     & <0,05 & Non-Trivial & Linear \\
    25    & 0,77  & 0,81  & 21    & <0,05 & Non-Trivial & Linear \\
    28    & 0,84  & 0,90  & 0     & <0,05 & Non-Trivial & Linear \\
    29    & 0,99  & 0,99  & 30    & <0,05 & Non-Trivial & Linear \\
    30    & 0,73  & 0,80  & 5     & <0,05 & Non-Trivial & Linear \\
    33    & 0,94  & 0,93  & 166   & <0,05 & Non-Trivial & Non-linear \\
    34    & 0,97  & 0,95  & 172   & <0,05 & Non-Trivial & Non-linear \\
    39    & 0,96  & 0,96  & 36    & <0,05 & Non-Trivial & Linear \\
    42    & 0,88  & 0,87  & 149   & 0,11  & Non-Trivial & Linear \\
    43    & 0,99  & 0,96  & 210   & <0,05 & Non-Trivial & Non-linear \\
    48    & 1,01  & 0,99  & 191   & <0,05 & Non-Trivial & Non-linear \\
    49    & 0,63  & 0,65  & 7     & <0,05 & Non-Trivial & Linear \\
    51    & 0,77  & 0,80  & 0     & <0,05 & Non-Trivial & Linear \\
    52    & 0,94  & 0,94  & 116   & 0,70  & Non-Trivial & Linear \\

    \hline
    \end{tabular}
\end{table}

One may assume either linearity or non-linearity from the results of our non-linearity test, but it is important to analyze the robustness of these results to both the presence of outliers and the impact of other data transformations.

\section{The impact of data transformations \label{sec:outliers}}

In this section, we aim to analyze the impact of outlier treatments on noise reduction, as intended, and on information loss, as an undesirable effect. We also study the influence of Box-Cox data transformations on the results of our cross-validated non-linearity test. Detection and treatment of outliers is an ongoing issue in data mining \citep{rousseeuw1987robust,hodge2004survey}. An outlier is an observation that appears to significantly deviate from other members of the sample in which it occurs \citep{grubbs1969procedures}. Outliers arise due to changes in systems, measurement errors or simply due to deviations from average activity. It is also important to note that an outlier may also be the most interesting part of the data. 

On the one hand, from the set of cash flow time-series labeled as trivial, some of them may be labeled as non-trivial after removing outliers as a way of noise reduction. On the other hand, from those data sets labeled as non-trivial, some of them may be labeled as trivial due to the information loss produced by the treatment. We here measure the effect of removing outliers on the prediction error using time-series cross validation for different thresholds of outlier replacement. For each data set, we progressively identify as outliers cash flow observations greater than 5, 4, and 3 times the standard deviation in a training set with the 80\% oldest observations. We replace outliers with a linear interpolation and proceed as detailed in Algorithm \ref{GeneralAlg} to cross-validate triviality and linearity. The results from this analysis are summarized in Table \ref{tab:outliers}.

By following this procedure, we identify data sets 5, 10, 17, 32, 44 and 54  (6 out of 28), initially labeled as trivial that, after outlier treatment, can be labeled as non-trivial due to noise reduction. Similarly, data sets 4 and 48 that were initially labeled as non-trivial can be labeled as trivial after outlier treatment due to information loss. If we measure noise reduction by the error reduction and information loss by the error increase, then we can assess the impact of outlier treatment. Following this approach, we obtained mixed results for non-trivial data sets after outlier treatment: an average noise reduction of 22\%, and an average information loss of 14\%. It is important to recall that unexpected observations are often the most interesting part of the data to predict, e.g., when the goal is to forecast unusual but genuine cash flows.

\begin{table}[t]
  \centering
  \tiny
  \caption{\label{tab:outliers}Results of the test for non-linearity after outlier treatment and Box-Cox transformation. Changes in labels are marked with $^*$. }
    \begin{tabular}{rllclc}
    \hline
          &       & \multicolumn{2}{c}{After outliers} & \multicolumn{2}{c}{After outliers and Box-Cox} \\
     \cline{3-4}
     \cline{5-6}
    \multicolumn{1}{c}{Id} & \multicolumn{1}{l}{Triviality} & \multicolumn{1}{l}{Linearity} & \multicolumn{1}{c}{Noise reduction} & \multicolumn{1}{l}{Linearity} & \multicolumn{1}{c}{Noise reduction} \\
    \hline
    1     & Non-Trivial & Linear & 0,00  & Non-linear$^*$ & -0,01 \\
    3     & Non-Trivial & Linear & 0,02  & Non-linear$^*$ & 0,00 \\
    5     & Non-Trivial & Non-linear & 0,40  & Non-linear & 0,41 \\
    7     & Non-Trivial & Linear & -0,10 & Linear & -0,13 \\
    9     & Non-Trivial & Linear & -0,04 & Linear & -0,04 \\
    10    & Non-Trivial & Non-linear & 0,46  & Non-linear & 0,47 \\
    13    & Non-Trivial & Linear & -0,18 & Linear & -0,21 \\
    14    & Non-Trivial & Linear & -0,05 & Linear & -0,07 \\
    16    & Non-Trivial & Linear & -0,18 & Linear & -0,17 \\
    17    & Non-Trivial & Non-linear$^*$ & 0,71  & Non-linear & 0,71 \\
    18    & Non-Trivial & Non-linear$^*$& -0,20 & Non-linear & -0,20 \\
    19    & Non-Trivial & Non-linear & -0,03 & Non-linear & -0,04 \\
    20    & Non-Trivial & Non-linear & -0,02 & Non-linear & -0,02 \\
    23    & Non-Trivial & Non-linear$^*$ & -0,22 & Non-linear & -0,22 \\
    24    & Non-Trivial & Linear & -0,20 & Linear & -0,06 \\
    25    & Non-Trivial & Non-linear$^*$ & -0,26 & Non-linear & -0,25 \\
    28    & Non-Trivial & Linear & -0,05 & Linear & -0,04 \\
    29    & Non-Trivial & Linear & 0,07  & Non-linear$^*$ & 0,00 \\
    30    & Non-Trivial & Linear & -0,06 & Linear & -0,04 \\
    32    & Non-Trivial & Non-linear & 0,18  & Non-linear & 0,21 \\
    33    & Non-Trivial & Linear$^*$ & -0,12 & Linear & -0,11 \\
    34    & Non-Trivial & Linear$^*$ & 0,12  & Linear & 0,09 \\
    39    & Non-Trivial & Non-linear$^*$ & -0,02 & Linear$^*$ & -0,01 \\
    42    & Non-Trivial & Linear & -0,23 & Linear & -0,14 \\
    43    & Non-Trivial & Non-linear & 0,04  & Non-linear & 0,03 \\
    44    & Non-Trivial & Non-linear$^*$ & 0,48  & Non-linear & 0,82 \\
    49    & Non-Trivial & Non-linear$^*$ & -0,56 & Non-linear & -0,61 \\
    51    & Non-Trivial & Linear & -0,03 & Linear & -0,03 \\
    52    & Non-Trivial & Linear & 0,01  & Linear & 0,03 \\
    54    & Non-Trivial & Linear$^*$ & 0,17  & Linear & 0,17 \\
    \hline
    \end{tabular}
\end{table}%

Non-linearity and outliers are closely linked. Indeed, \cite{castle2012automatic} hypothesized that non-linear functions can align with outliers, causing functions to be considered relevant spuriously, which can be detrimental for generalizing and forecasting. If this hypothesis is correct, the relative forecasting ability of a linear model in comparison to a non-linear model would increase as the presence of outliers in a training set is reduced. From the set of time-series finally labeled as non-trivial, data sets 33, 34 and 54, initially labeled as non-linear changed their labels to linear. Surprisingly, data sets 17, 18, 23, 25, 39, 44 and 49 (7 out of 30), could be labeled as non-linear after outlier treatment. Except for data sets 17 and 44, in all cases there was information loss, i.e., error increase, suggesting that non-linear models can deal better with information loss.

We also considered a Box-Cox transformation to analyze if this kind of data transformation may influence the results from our cross-validated non-linearity test. From the set of non-trivial data sets we compare linearity labels, first, after outlier treatment, and second, after outlier treatment and Box-Cox transformation as described in equation (\ref{eq:boxcox}). In addition, we compare information loss computed as the difference between the sum of errors of the linear and non-linear forecasting models before and after the outlier treatment. A positive value means noise reduction or error reduction while a negative value means information loss or error increase. These results show that our cross-validated non-linearity test outputs similar results after Box-Cox transformation since the change in labels were produced in data sets with similar linear and non-linear model performance. A summary of the results of this section is shown in Table \ref{tab:sumresults}.

The high number of trivial data sets may be caused by the general inherent randomness of cash flows. Outlier treatment produced a small improvement in non-triviality but also an outstanding increase in non-linearity. Finally, Box-Cox data transformation yielded similar results but with better results for non-linear models. Thus, we conclude that: (i) common data transformations had little impact on our time-series in terms of linearity; and (ii) outlier treatment and Box-Cox transformation were unable to transform non-linear into linear cash flows.

\begin{table}[!htb]
  \centering
  \caption{\label{tab:sumresults} Number of time-series data sets and their labels after transformation. OT=Outlier treatment; DT=Data transformation.}
    \begin{tabular}{lccc}
    \hline
	Label & Raw data & After OT & After OT and DT \\
    \hline
	Trivial & 28 & 24 & 24 \\
    Non-trivial & 26 & 30 & 30 \\
    \hspace{5mm}-Linear & 20 & 17 & 15 \\
    \hspace{5mm}-Non-linear & 6 & 13 & 15 \\
    \hline
    \end{tabular}
\end{table}%

\section{Concluding remarks\label{sec:conclusions}}

Small and medium companies are the most common type of companies in Europe contributing to a high percentage of all enterprises, value added and employment. In this paper, we provide a complete empirical study of the statistical properties of daily cash flows based on 54 real-world cash flow time-series from this kind of companies. To the best of our knowledge, this study is the most comprehensive empirical study on daily cash flow so far in terms of the range of statistical properties considered, and also on both the number and the length of the data sets that we make available online. We particularly focus on the implications for forecasting due to its key role in cash management.

\subsection{Summary of findings}

Our results show that the extended hypothesis of normal, stationary and uncorrelated cash flows is hardly present in our cash flow data set. We do not claim that these results can be generalized to all small and medium companies. However, we conclude that the standard assumptions of normality, stationarity and uncorrelatedness that have been extensively used in cash management literature must be verified before the deployment of any cash management model based on them. We also highlight that common solutions to non-normality and non-stationarity such as data transformation and differencing produce little benefit when applied to our data sets, with the risk of loosing important information on extreme cash flows. Alternative and more complex data transformations are nevertheless an option to consider in further research to achieve Gaussian cash flows.

In an attempt to discover the attributes of actual-world cash flows, we also studied the presence of non-linearity. To this end, we proposed a new simple test for non-linearity with two main advantages in comparison to alternative approaches. First, our test does not assume any non-linear function. Second, it is based on time-series cross validation to increase robustness and avoid overfitting. It is important to note that our cross-validated definition of non-linearity depends on the alternative models considered, one linear and another one non-linear.

Our cross-validated non-linearity test labeled as either trivial, linear or non-linear our cash flow data set after outlier treatment resulting in an important increase the number of data sets labeled as non-linear. After both outlier treatment and Box-Cox transformation, linearity could not be achieved and non-linear models showed more robust. The application of our test to provide further evidence on these topics when using alternative cash flow data sets represents a natural extension of our work.

\subsection{Implications}

Our results raise questions about two common assumptions in cash flow time-series since we found that: (i) the usual assumption of normality, absence of correlation and stationarity is hardly present; and (ii) common data transformations such as outlier treatment and Box-Cox transformation have little impact on normality and linearity. Contrary to the rather common assumption in the literature, these results imply that neither it is always possible to achieve a Gaussian, white-noise and linear time-series through data transformation nor it is always desirable due to information loss. Thus, linear models should be considered as an initial step towards more realistic ones which are better adapted to real cash flow situations. The results from our cross-validated test for non-linearity suggest that non-linear models represent a justifiable alternative for time-series forecasting. Moreover, since our test is both model and outlier dependent, a promising line of future work would be the integration of outlier treatment in the test itself in an attempt to assess noise reduction or information loss.

As a result, we claim that a number of preliminary steps are necessary in cash flow forecasting before model selection: (i) statistical summary including normality, correlation and stationarity; (ii) impact of data transformations such as outlier treatment and Box-Cox transformation; (iii) non-linearity test to determine the type of model which is expected to deliver a better performance. Finally, this process is not limited to daily cash flow, since it can also be applied to any other time-series data set when cross-validation is required. 




\bibliography{bibliography}{}

\begin{thebibliography}{}

\bibitem[\protect\astroncite{Baccarin}{2009}]{baccarin2009optimal}
Baccarin, S. (2009).
\newblock Optimal impulse control for a multidimensional cash management system
  with generalized cost functions.
\newblock {\em European Journal of Operational Research}, 196(1):198--206.

\bibitem[\protect\astroncite{Baumol}{1952}]{baumol1952transactions}
Baumol, W.~J. (1952).
\newblock The transactions demand for cash: An inventory theoretic approach.
\newblock {\em The Quarterly Journal of Economics}, 66(4):545--556.

\bibitem[\protect\astroncite{Booth et~al.}{2014}]{booth2014automated}
Booth, A., Gerding, E., and Mcgroarty, F. (2014).
\newblock Automated trading with performance weighted random forests and
  seasonality.
\newblock {\em Expert Systems with Applications}, 41(8):3651--3661.

\bibitem[\protect\astroncite{Box and Cox}{1964}]{box1964analysis}
Box, G.~E. and Cox, D.~R. (1964).
\newblock An analysis of transformations.
\newblock {\em Journal of the Royal Statistical Society. Series B
  (Methodological)}, pages 211--252.

\bibitem[\protect\astroncite{Box and Jenkins}{1976}]{box1976time}
Box, G.~E. and Jenkins, G.~M. (1976).
\newblock {\em Time series analysis: forecasting and control, revised ed}.
\newblock Holden-Day.

\bibitem[\protect\astroncite{Breiman}{2001}]{breiman2001random}
Breiman, L. (2001).
\newblock Random forests.
\newblock {\em Machine learning}, 45(1):5--32.

\bibitem[\protect\astroncite{Broomhead and
  Lowe}{1988}]{broomhead1988multivariable}
Broomhead, D. and Lowe, D. (1988).
\newblock Multivariable functional interpolation and adaptive networks.
\newblock {\em Complex Systems}, 2(3):321--355.

\bibitem[\protect\astroncite{Castle and Hendry}{2010}]{castle2010low}
Castle, J.~L. and Hendry, D.~F. (2010).
\newblock A low-dimension portmanteau test for non-linearity.
\newblock {\em Journal of Econometrics}, 158(2):231--245.

\bibitem[\protect\astroncite{Castle and Hendry}{2012}]{castle2012automatic}
Castle, J.~L. and Hendry, D.~F. (2012).
\newblock Automatic selection for non-linear models.
\newblock In {\em System Identification, Environmental Modelling, and Control
  System Design}, pages 229--250. Springer.

\bibitem[\protect\astroncite{Cleveland et~al.}{1990}]{cleveland1990stl}
Cleveland, R.~B., Cleveland, W.~S., McRae, J.~E., and Terpenning, I. (1990).
\newblock Stl: A seasonal-trend decomposition procedure based on loess.
\newblock {\em Journal of Official Statistics}, 6(1):3--73.

\bibitem[\protect\astroncite{Constantinides and
  Richard}{1978}]{constantinides1978existence}
Constantinides, G.~M. and Richard, S.~F. (1978).
\newblock Existence of optimal simple policies for discounted-cost inventory
  and cash management in continuous time.
\newblock {\em Operations Research}, 26(4):620--636.

\bibitem[\protect\astroncite{da~Costa~Moraes et~al.}{2015}]{da2015stochastic}
da~Costa~Moraes, M.~B., Nagano, M.~S., and Sobreiro, V.~A. (2015).
\newblock Stochastic cash flow management models: A literature review since the
  1980s.
\newblock In {\em Decision Models in Engineering and Management}, pages 11--28.
  Springer International Publishing.

\bibitem[\protect\astroncite{Dietterich}{2000}]{dietterich2000ensemble}
Dietterich, T.~G. (2000).
\newblock Ensemble methods in machine learning.
\newblock In {\em Multiple classifier systems}, pages 1--15. Springer.

\bibitem[\protect\astroncite{Emery}{1981}]{emery1981some}
Emery, G.~W. (1981).
\newblock Some empirical evidence on the properties of daily cash flow.
\newblock {\em Financial management}, 10(1):21--28.

\bibitem[\protect\astroncite{Franses and
  Van~Dijk}{2005}]{franses2005forecasting}
Franses, P.~H. and Van~Dijk, D. (2005).
\newblock The forecasting performance of various models for seasonality and
  nonlinearity for quarterly industrial production.
\newblock {\em International Journal of Forecasting}, 21(1):87--102.

\bibitem[\protect\astroncite{Gormley and Meade}{2007}]{gormley2007utility}
Gormley, F.~M. and Meade, N. (2007).
\newblock The utility of cash flow forecasts in the management of corporate
  cash balances.
\newblock {\em European Journal of Operational Research}, 182(2):923--935.

\bibitem[\protect\astroncite{Gregory}{1976}]{gregory1976cash}
Gregory, G. (1976).
\newblock Cash flow models: a review.
\newblock {\em Omega}, 4(6):643--656.

\bibitem[\protect\astroncite{Grubbs}{1969}]{grubbs1969procedures}
Grubbs, F.~E. (1969).
\newblock Procedures for detecting outlying observations in samples.
\newblock {\em Technometrics}, 11(1):1--21.

\bibitem[\protect\astroncite{Ho}{1998}]{ho1998random}
Ho, T.~K. (1998).
\newblock The random subspace method for constructing decision forests.
\newblock {\em IEEE transactions on pattern analysis and machine intelligence},
  20(8):832--844.

\bibitem[\protect\astroncite{Hodge and Austin}{2004}]{hodge2004survey}
Hodge, V.~J. and Austin, J. (2004).
\newblock A survey of outlier detection methodologies.
\newblock {\em Artificial intelligence review}, 22(2):85--126.

\bibitem[\protect\astroncite{Homonoff and Mullins}{1975}]{homonoff1975cash}
Homonoff, R. and Mullins, D.~W. (1975).
\newblock {\em Cash management: an inventory control limit approach}.
\newblock Lexington Books.

\bibitem[\protect\astroncite{Hornik et~al.}{1989}]{hornik1989multilayer}
Hornik, K., Stinchcombe, M., and White, H. (1989).
\newblock Multilayer feedforward networks are universal approximators.
\newblock {\em Neural networks}, 2(5):359--366.

\bibitem[\protect\astroncite{Hyndman and
  Athanasopoulos}{2013}]{hyndman2013forecasting}
Hyndman, R. and Athanasopoulos, G. (2013).
\newblock Forecasting: principles and practice.
\newblock \url{http://otexts.org/fpp/}.
\newblock Last accessed: 2017-05-01.

\bibitem[\protect\astroncite{Kantz and Schreiber}{2004}]{kantz2004nonlinear}
Kantz, H. and Schreiber, T. (2004).
\newblock {\em Nonlinear time series analysis}, volume~7.
\newblock Cambridge University Press.

\bibitem[\protect\astroncite{Keenan}{1985}]{keenan1985tukey}
Keenan, D.~M. (1985).
\newblock A tukey nonadditivity-type test for time series nonlinearity.
\newblock {\em Biometrika}, 72(1):39--44.

\bibitem[\protect\astroncite{Lee et~al.}{1993}]{lee1993testing}
Lee, T.-H., White, H., and Granger, C.~W. (1993).
\newblock Testing for neglected nonlinearity in time series models: A
  comparison of neural network methods and alternative tests.
\newblock {\em Journal of Econometrics}, 56(3):269--290.

\bibitem[\protect\astroncite{Lilliefors}{1967}]{lilliefors1967kolmogorov}
Lilliefors, H.~W. (1967).
\newblock On the kolmogorov-smirnov test for normality with mean and variance
  unknown.
\newblock {\em Journal of the American Statistical Association},
  62(318):399--402.

\bibitem[\protect\astroncite{Ljung and Box}{1978}]{ljung1978measure}
Ljung, G.~M. and Box, G.~E. (1978).
\newblock On a measure of lack of fit in time series models.
\newblock {\em Biometrika}, 65(2):297--303.

\bibitem[\protect\astroncite{Makridakis
  et~al.}{2008}]{makridakis2008forecasting}
Makridakis, S., Wheelwright, S.~C., and Hyndman, R.~J. (2008).
\newblock {\em Forecasting methods and applications}.
\newblock John Wiley \& Sons.

\bibitem[\protect\astroncite{Miller and Orr}{1966}]{miller1966model}
Miller, M.~H. and Orr, D. (1966).
\newblock A model of the demand for money by firms.
\newblock {\em The Quarterly journal of economics}, 80(3):413--435.

\bibitem[\protect\astroncite{Miller and Stone}{1985}]{miller1985daily}
Miller, T.~W. and Stone, B.~K. (1985).
\newblock Daily cash forecasting and seasonal resolution: Alternative models
  and techniques for using the distribution approach.
\newblock {\em Journal of Financial and Quantitative Analysis}, 20(3):335--351.

\bibitem[\protect\astroncite{Muller et~al.}{2015}]{muller2015annual}
Muller, P., Caliandro, C., Peycheva, V., Gagliardi, D., Marzocchi, C.,
  Ramlogan, R., and Cox, D. (2015).
\newblock Annual report on european {SMEs}.

\bibitem[\protect\astroncite{Nedovi{\'c} and
  Deved{\v{z}}i{\'c}}{2002}]{nedovic2002expert}
Nedovi{\'c}, L. and Deved{\v{z}}i{\'c}, V. (2002).
\newblock Expert systems in finance—a cross-section of the field.
\newblock {\em Expert Systems with Applications}, 23(1):49--66.

\bibitem[\protect\astroncite{Penttinen}{1991}]{penttinen1991myopic}
Penttinen, M.~J. (1991).
\newblock Myopic and stationary solutions for stochastic cash balance problems.
\newblock {\em European Journal of Operational Research}, 52(2):155--166.

\bibitem[\protect\astroncite{Pindado and Vico}{1996}]{pindado1996evidencia}
Pindado, J. and Vico, J. (1996).
\newblock Evidencia emp{\'\i}rica sobre los flujos de caja. un nuevo enfoque en
  su tratamiento.
\newblock {\em Revista Espa{\~n}ola de Financiaci{\'o}n y Contabilidad},
  25(87):497--517.

\bibitem[\protect\astroncite{Premachandra}{2004}]{premachandra2004diffusion}
Premachandra, I. (2004).
\newblock A diffusion approximation model for managing cash in firms: An
  alternative approach to the miller--orr model.
\newblock {\em European Journal of Operational Research}, 157(1):218--226.

\bibitem[\protect\astroncite{Ramsey}{1969}]{ramsey1969tests}
Ramsey, J.~B. (1969).
\newblock Tests for specification errors in classical linear least-squares
  regression analysis.
\newblock {\em Journal of the Royal Statistical Society. Series B
  (Methodological)}, 31(2):350--371.

\bibitem[\protect\astroncite{Ross et~al.}{2002}]{ross2002fundamentals}
Ross, S.~A., Westerfield, R., and Jordan, B.~D. (2002).
\newblock {\em Fundamentals of corporate finance}.
\newblock McGraw-Hill, sixth edition.

\bibitem[\protect\astroncite{Rousseeuw and Leroy}{1987}]{rousseeuw1987robust}
Rousseeuw, P.~J. and Leroy, A.~M. (1987).
\newblock {\em Robust regression and outlier detection}, volume 589.
\newblock John Wiley \& Sons.

\bibitem[\protect\astroncite{Royston}{1982}]{royston1982extension}
Royston, J. (1982).
\newblock An extension of shapiro and wilk's w test for normality to large
  samples.
\newblock {\em Applied Statistics}, 31(2):115--124.

\bibitem[\protect\astroncite{Salas-Molina et~al.}{2017}]{salas2017empowering}
Salas-Molina, F., Martin, F.~J., Rodr{\'\i}guez-Aguilar, J.~A., Serr{\`a}, J.,
  and Arcos, J.~L. (2017).
\newblock Empowering cash managers to achieve cost savings by improving
  predictive accuracy.
\newblock {\em International Journal of Forecasting}, 33(2):403--415.

\bibitem[\protect\astroncite{Srinivasan and
  Kim}{1986}]{srinivasan1986deterministic}
Srinivasan, V. and Kim, Y.~H. (1986).
\newblock Deterministic cash flow management: state of the art and research
  directions.
\newblock {\em Omega}, 14(2):145--166.

\bibitem[\protect\astroncite{Stone and Miller}{1987}]{stone1987daily}
Stone, B.~K. and Miller, T.~W. (1987).
\newblock Daily cash forecasting with multiplicative models of cash flow
  patterns.
\newblock {\em Financial Management}, 16(4):45--54.

\bibitem[\protect\astroncite{Stone and Wood}{1977}]{stone1977daily}
Stone, B.~K. and Wood, R.~A. (1977).
\newblock Daily cash forecasting: a simple method for implementing the
  distribution approach.
\newblock {\em Financial Management}, 6(3):40--50.

\bibitem[\protect\astroncite{Venables and Ripley}{2013}]{venables2013modern}
Venables, W.~N. and Ripley, B.~D. (2013).
\newblock {\em Modern applied statistics with S-PLUS}.
\newblock Springer Science \& Business Media.

\bibitem[\protect\astroncite{Wilcoxon et~al.}{1970}]{wilcoxon1970critical}
Wilcoxon, F., Katti, S., and Wilcox, R.~A. (1970).
\newblock Critical values and probability levels for the wilcoxon rank sum test
  and the wilcoxon signed rank test.
\newblock {\em Selected tables in mathematical statistics}, 1:171--259.

\bibitem[\protect\astroncite{Zagorecki}{2015}]{zagorecki2015prediction}
Zagorecki, A. (2015).
\newblock Prediction of methane outbreaks in coal mines from multivariate time
  series using random forest.
\newblock In {\em Rough Sets, Fuzzy Sets, Data Mining, and Granular Computing},
  pages 494--500. Springer.

\bibitem[\protect\astroncite{Zhang et~al.}{1998}]{zhang1998forecasting}
Zhang, G., Patuwo, B.~E., and Hu, M.~Y. (1998).
\newblock Forecasting with artificial neural networks:: The state of the art.
\newblock {\em International Journal of Forecasting}, 14(1):35--62.

\end{thebibliography}

\end{document}